# ON THE PROCEDURE OF EXAMINATION OF INSTABILITY OF THE SHOCK FRONT RELATIVELY ACOUSTO-GRAVITATIONAL PERTURBATIONS

**V.V. Lyahov, V.M. Neshchadim** [(1,2)]

Institute of Ionosphere, Kamenskoe plato, 050020 Almaty, Kazakhstan

## INTRODUCTION

It is known that quick generation of large energy in small field of space forms shock wave. Instantaneous point source model may be used to solve many problems. Our concern is atmosphere response to point land and underground, industrial and military explosions of high power. The problem of study of amplitude-frequency characteristics of atmospheric disturbance induced by earthquakes is within this field of interest.

The shock wave is propagating relative to the medium at a supersonic speed until pressure on its front does not drop to the value of pressure of the undisturbed medium. In this case, velocity of shock wave reduces to sound velocity in a given point of medium, and the shock wave degenerates into a usual acoustic wave. The shock front under certain conditions can be unstable relative to various perturbations. In "Theoretical Physics" of L.D.Landau, E.M.Lifshits [1] there is a deduction of the dispersing equation for unstable stability of shock front relative to sound perturbations. We examine more a general case of unstable stability of front relative to acousto-gravitational perturbations. The dispersing equation [1] does not contain a gravitational branch of oscillations that is interesting for us; therefore, we will derive a dispersing equation for acousto-gravitational modes of unstable stability of shock front.

## STATEMENT OF THE PROBLEM

Let's find relations between velocity modifications and thermodynamic quantities at small perturbation of the homogeneous stream of gas (air). The atmosphere is described by the system of hydrodynamic equations:

$$\rho \frac{\partial \vec{v}}{\partial t} + \rho(\vec{v}\nabla\vec{v}) = -\nabla p + \rho\vec{g},$$

$$\frac{\partial \rho}{\partial t} + \nabla(\rho\vec{v}) = 0, \qquad (1)$$

$$\frac{\partial s}{\partial t} + \vec{v}\nabla s = 0.$$

Where: $\rho$ - medium density, $\vec{v}$ - stream velocity, $p$ - pressure, $\vec{g}$ - acceleration of gravity, $s$ - entropy.

---
[(1,2)] mail: v_lyahov@rambler.ru, ne_dim@bk.ru

## PROBLEM SOLUTION

We solve the problem using perturbation technique:

$$\vec{v}(\vec{r},t) = \vec{v}_0 + \delta\vec{v}(\vec{r},t),$$
$$p(\vec{r},t) = p_0(\vec{r}) + \delta p(\vec{r},t),$$
$$\rho(\vec{r},t) = \rho_0(\vec{r}) + \delta\rho(\vec{r},t), \quad (2)$$
$$s(\vec{r},t) = s_0(\vec{r}) + \delta s(\vec{r},t).$$

Perturbations of all quantities are small (much less than basic value). Substituting decompositions (2) in system (1), we will derive equations for perturbations:

$$\frac{\partial \delta\vec{v}}{\partial t} + (\vec{v}_0 \nabla)\delta\vec{v} + \frac{\nabla \delta p}{\rho_0} - \frac{\vec{g}}{\rho_0}\delta\rho = 0,$$

$$\frac{\partial \delta p}{\partial t} + \vec{v}_0 \nabla \delta p + \rho_0 c^2 \nabla \delta\vec{v} = 0, \quad (3)$$

$$\frac{\partial \delta s}{\partial t} + \vec{v}_0 \nabla \delta s = 0.$$

At derivation of the equation of continuity (the second equation in the last system) $\delta p = \left(\frac{\partial p}{\partial \rho}\right)_s \delta\rho$ and determination of sound velocity $c = \sqrt{\left(\frac{\partial p}{\partial \rho}\right)_s}$ have been used.

Expression of perturbations in the form of plane waves

$$\delta\vec{v}(\vec{r},t) = \delta\vec{v}(\omega,\vec{k})\exp[i(-\omega t + \vec{k}\vec{r})],$$
$$\delta p(\vec{r},t) = \delta p(\omega,\vec{k})\exp[i(-\omega t + \vec{k}\vec{r})],$$
$$\delta\rho(\vec{r},t) = \delta\rho(\omega,\vec{k})\exp[i(-\omega t + \vec{k}\vec{r})], \quad (4)$$
$$\delta s(\vec{r},t) = \delta s(\omega,\vec{k})\exp[i(-\omega t + \vec{k}\vec{r})],$$

makes it possible to derive from system (3) the following algebraic equations for Fouriers-amplitudes:

$$(\vec{v}_0\vec{k} - \omega)\delta\vec{v} + \frac{\vec{k}}{\rho_0}\delta p + i\frac{\vec{g}}{\rho_0}\delta\rho = 0,$$

$$(\vec{v}_0\vec{k} - \omega)\delta s = 0, \quad (5)$$

$$(\vec{v}_0\vec{k} - \omega)\delta p + \rho_0 c^2 \vec{k}\delta\vec{v} = 0.$$

The first equation of system (5) differs from the analogous equation of the system derived in [1]. Supposing $\vec{g} = 0$, i.e. gravity is to taken into account, than system (5) becomes identical to system [1]. It is the third addend of the first equation of system (5) that bears information on a gravitational branch of acousto-gravitational oscillations.

The system of equations (5) admits two types of perturbations. In the first of them (so-called entropy-vortex waves) the following relations between perturbed quantities are realized:

$$\omega = \vec{v}_0\vec{k},\ \delta s \neq 0,\ \delta p = -i\frac{\vec{g}}{\vec{k}}\delta\rho,\ \delta\rho = \frac{1}{c^2}\delta p + \left(\frac{\partial\rho}{\partial s}\right)_p \delta s,\ \vec{k}\delta\vec{v} = 0. \qquad (6)$$

In the second type:

$$(\omega - \vec{v}_0\vec{k})^2 = c^2 k^2,\ \delta s = 0,\ \delta p = c^2\delta\rho,\ (\omega - \vec{v}_0\vec{k})\delta p = \rho_0 c^2 \vec{k}\delta\vec{v},\ [\vec{k}\delta\vec{v}] = \frac{\delta\rho}{\rho_0}[\vec{k}\vec{g}]. \qquad (7)$$

These are acousto-gravitational waves with the frequency shifted due to Doppler Effect.

We consider a flat shock wave. Fig. 1 shows a system of coordinates related to this wav. The shock wave is perpendicular to Z-axes. The origin of coordinates is on a discontinuity surface. The gas stream flows through the discontinuity surface from the left to the right on the Z-axis.

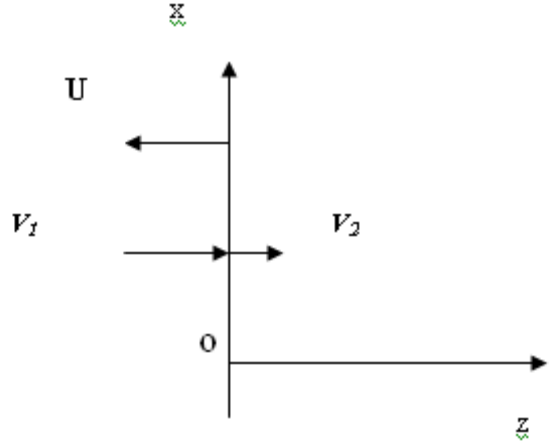

Fig.1. The frame of axis related to a shock wave

We consider that the discontinuity surface undergoes a small perturbation in the form of wave:

$$\xi = \xi_0 \exp[i(-\omega t + k_x x)]. \qquad (8)$$

This surface perturbation induces perturbation of gas flow behind the shock wave in $z > 0$ domain. Flow before disruption $z < 0$ does not undergo perturbation due to it's supersonic velocity.

Perturbation of gas flow behind the shock front consists of entropy-vortex waves (EVW) and acousto-gravitational waves (AGW):

$$\delta v_{x,y} = \delta v_{x,y}^{(evw)} + \delta v_{x,y}^{(AGW)},\ \delta V = \delta V^{(evw)} + \delta V^{(AGW)},\ \delta p = \delta p^{(evw)} + \delta p^{(AGW)}. \qquad (9)$$

Here, $V = \frac{1}{\rho}$ is specific volume

Using formulas (6) and (7), we will combine 9 quantities:

$$\xi, \delta v_x^{(evw)}, \delta v_x^{(AGW)}, \delta v_y^{(evw)}, \delta v_y^{(AGW)}, \delta p^{(evw)}, \delta p^{(AGW)}, \delta V^{(evw)}, \delta V^{(AGW)}.$$

In our statement of the problem, $\vec{v}_0 = \vec{v}_2$, we will consider plane waves of perturbations (4) as two-dimensional, i.e. $k^2 = k_x^2 + k_z^2$.

For entropy-vortex wave it follows from (6) that $\vec{k}\vec{v}_2 = \omega$, i.e. $k_z = \dfrac{\omega}{v_2}$, and $\vec{k}\delta\vec{v}^{(evw)} = 0$, therefore, we derive the final equation:

$$\frac{\omega}{v_2}\delta v_z^{(evw)} + k_x \delta v_x^{(evw)} = 0. \tag{10}$$

From the third equity of expression (6) we will derive:

$$\delta p^{(evw)} = -i\frac{\vec{g}}{\vec{k}}\delta\rho^{(evw)} = -i\frac{\vec{g}}{\vec{k}}\delta\left(\frac{1}{V^{(evw)}}\right) = i\frac{\vec{g}}{\vec{k}}\frac{1}{V_2^2}\delta V^{(evw)}. \tag{11}$$

For AGW it follows from (7) that $\delta p = c^2 \delta\rho = c^2 \delta(\frac{1}{V}) = -\dfrac{c^2}{V^2}\delta V$ and the final equation is:

$$\delta p^{(AGW)} = -\left(\frac{c_2}{V_2}\right)^2 \delta V^{(AGW)}. \tag{12}$$

Let's convert the fourth formula of expression (7) by scalar multiplication of both parts of the equity by vector $\vec{k}$ and take into account the first equity (7):

$$(\omega - v_2 k_z)\delta v_z^{(AGW)} = V_2 k_z \delta p^{(AGW)}, \tag{13}$$

$$(\omega - v_2 k_z)\delta v_x^{(AGW)} = V_2 k_x \delta p^{(AGW)}. \tag{14}$$

The remaining relations for perturbed quantities are derived from boundary conditions on the discontinuity surface perturbed by periodic oscillations (8) and therefore coincide with the relations derived in [1] previously to the accuracy similar to that of the case under consideration $\delta p^{(evw)} \neq 0$.

$$\delta v_x^{(evw)} + \delta v_x^{(AGW)} = ik\xi(v_1 - v_2), \tag{15}$$

$$\delta v_z^{(evw)} + \delta v_z^{(AVW)} = \frac{v_2 - v_1}{2}\left(\frac{\delta p^{(evw)} + \delta p^{(AGW)}}{p_2 - p_1} - \frac{\delta V^{(evw)} + \delta V^{(AGW)}}{V_1 - V_2}\right), \tag{16}$$

$$\delta p^{(evw)} + \delta p^{(AGW)} = \frac{dp_2}{dV_2}(\delta V^{(evw)} + \delta V^{(AGW)}), \tag{17}$$

$$i\frac{2\omega}{v_1}\xi = \frac{\delta p^{(evw)} + \delta p^{(AGW)}}{p_2 - p_1} + \frac{\delta V^{(evw)} + \delta V^{(AGW)}}{V_1 - V_2}. \tag{18}$$

Equations (10), (11), (12), (13), (14), (15), (16), (17), (18) represent system of 9 linear algebraic equations for 9 quantities:

$\xi, \delta v_x^{(evw)}, \delta v_x^{(AGW)}, \delta v_y^{(evw)}, \delta v_y^{(AGW)}, \delta p^{(evw)}, \delta p^{(AGW)}, \delta V^{(evw)}, \delta V^{(AGW)}$.

Nontrivial solution of this system expressed by equality of the determinant of this system to null gives the dispersing equation:

$$\left| \Lambda_{ij}(\omega, k_x, k_z) \right| = 0, \tag{19}$$

where

$$\Lambda_{16} = \frac{\omega}{v_2}, \Lambda_{18} = k_x,$$

$$\Lambda_{23} = -V_2 k_z, \Lambda_{27} = \omega - v_2 k_z,$$

$$\Lambda_{33} = -V_2 k_x, \Lambda_{39} = \omega - v_2 k_z,$$

$$\Lambda_{41} = -ik(v_1 - v_2), \Lambda_{48} = 1, \Lambda_{49} = 1,$$

$$\Lambda_{52} = -\frac{v_2 - v_1}{2(p_2 - p_1)}, \Lambda_{53} = -\frac{v_2 - v_1}{2(p_2 - p_1)}, \Lambda_{54} = \frac{v_2 - v_1}{2(V_1 - V_2)}, \Lambda_{55} = \frac{v_2 - v_1}{2(V_1 - V_2)}, \Lambda_{56} = 1, \Lambda_{57} = 1,$$

$$\Lambda_{63} = 1, \Lambda_{65} = \left(\frac{c_2}{V_2}\right)^2, \tag{20}$$

$$\Lambda_{72} = 1, \Lambda_{74} = \frac{\vec{g}}{k} \frac{1}{V_2^2},$$

$$\Lambda_{82} = 1, \Lambda_{83} = 1, \Lambda_{84} = -\frac{dp_2}{dV_2}, \Lambda_{85} = -\frac{dp_2}{dV_2},$$

$$\Lambda_{91} = i\frac{2\omega}{v_1}, \Lambda_{92} = -\frac{1}{p_2 - p_1}, \Lambda_{93} = -\frac{1}{p_2 - p_1}, \Lambda_{94} = -\frac{1}{V_1 - V_2}, \Lambda_{95} = -\frac{1}{V_1 - V_2}.$$

The remaining components are $\Lambda_{ij} = 0$.

## CONCLUSION

The derived dispersing equation (19) makes it possible to study instability stability of the shock front not only relative to sound vibrations but also to explore the gravitational branch of acousto-gravitational waves. That is, within the model under study there appeared a possibility to study the interval of frequencies within which wave disturbances of the atmosphere induced by shock waves from land-based sources are excited.

## REFERENCES

Landau L.D., Lifshits E.M. Theoretical Physics, Vol. 6, Hydrodynamics, Moscow: Science, 1986, p. 472.